\documentclass[prb,twocolumn,showpacs,superscriptaddress]{revtex4}
\usepackage{graphicx}
\usepackage{dcolumn}
\usepackage{bm}

\begin{document}
\title{Slave-boson based configuration-interaction approach
for the Hubbard model}
\author{G. Seibold}
\affiliation{Institut f\"ur Physik, BTU Cottbus, PBox 101344,
         03013 Cottbus, Germany}

\begin{abstract}
Based on the Kotliar-Ruckenstein slave-boson scheme we develop 
a configuration-interaction (CI) approach which is suitable to improve
the energy of symmetry-broken saddle-point  solutions.
The theory is applied to spin-polaron states in the Hubbard model
and compared with analogous results obtained within the
Hartree-Fock approximation. 
In addition we show that within the infinite ${\cal D}$ prescription of 
the Gutzwiller method a CI approach does not improve the variational result 
since in the thermodynamic limit matrix elements between different 
inhomogeneous states vanish due to an 'orthogonality catastrophe'.


\end{abstract}
\maketitle

\section{Introduction}
The Gutzwiller Ansatz is a variational
wave function for correlated electronic models with purely
local interaction.~\cite{GUTZ1,GUTZ2}
The basic idea to treat these Hubbard-type hamiltonians is 
to partially project 
out configurations with
doubly-occupied sites from the Fermi sea in order to optimize
the contributions from kinetic and potential energy.
As a consequence, in contrast to the conventional Hartree-Fock (HF) theory,
the Gutzwiller wave function captures correlation effects
like the band narrowing already on the variational level.
However, the exact evaluation
of the ground state energy within the Gutzwiller wave function
is fairly difficult and
up to now has only been achieved in one and infinite dimensions.~\cite{METZNER}
In the latter case the solution is equivalent to the
so-called Gutzwiller approximation (GA) which has been
applied to describe a variety of finite dimensional systems ranging from
the properties of normal $^3$He (cf. Ref.~\onlinecite{VOLLHARDT})
to the stripe phase of high-T$_c$ cuprates.~\cite{GOETZ,LOR}

The GA in its original formulation was restricted to homogeneous
paramagnetic systems and only later on generalized to arbitrary
Slater determinants by Gebhard~\cite{GEBHARD} and, more recently, by
Attaccalite and Fabrizio.~\cite{michele}
The same energy functional
was obtained from the Kotliar-Ruckenstein (KR) slave-boson formulation
of the Hubbard model when the bosons are replaced by their
mean-values.~\cite{KR}
Unconstrained minimization of the KR (or Gebhards) energy functional
on finite clusters in general yields inhomogeneous solutions
which break translational and spin-rotational invariance.~\cite{sei98,sei982}
This approach has been used for the investigation of 
electronic inhomogeneities,
such as stripes and checkerboards \cite{lor03,sei04,sei07},
in the context of high-T$_c$ superconductors.

Incorporation of fluctuations in the frame of the time-dependent
Gutzwiller approximation tends to restore the original symmetry of 
the system.~\cite{goe01} An alternative would be the construction of
a wave-function which is a linear superposition of equivalent
symmetry-broken states. In case of stripe states \cite{lor03,sei04} 
one could e.g. envisage a superposition of solutions which are translated
perpendicular to the stripe direction and also the corresponding
solutions which are rotated by $90$ degrees.
In case of the unrestricted Hartree-Fock approximation such a 
configuration-interaction (CI) method  
has been proposed in Ref. \onlinecite{louis99} and applied to the
case of stripe textures in Ref. \onlinecite{louis01}.

The present paper investigates the possibility wether an improvement
of the inhomogeneous Gutzwiller approximation is possible within an
analogous framework. 
In Sec. II we evaluate the matrix elements
of the Hubbard hamiltonian between different inhomogeneous solutions 
obtained from the saddle-point approximation of the KR slave-boson
scheme.~\cite{KR} Based on these results we construct a 
a CI method which in Sec. III is applied to spin polaron
states. We compare ground state energies with exact diagonalization results
and for larger lattices evaluate the dispersion relation of the
spin polaron states which can be compared with analogous solutions obtained
in the tJ-model. In this context we also compare our results
with angle-resolved photoemission (ARPES) experiments on Sr$_2$CuO$_2$Cl$_2$.
 
In appendix \ref{secga} it is shown that the infinite ${\cal D}$ prescription
of the Gutzwiller approximation \cite{GEBHARD} cannot be used for 
an analogous construction of a CI approach. The reason is that 
in the thermodynamic limit this scheme leads to an 'orthogonality 
catastrophe' \cite{and67} so that energy corrections and the dispersion of
quasiparticles vanish. 

\section{Model and Formalism}
Our investigations are based on the one-band Hubbard model
\begin{equation}\label{HM}
H=\sum_{ij,\sigma}t_{ij}c_{i,\sigma}^{\dagger}c_{j,\sigma}
+ U\sum_{i} n_{i,\uparrow}n_{i,\downarrow}
\end{equation}
where $c_{i,\sigma}^{(\dagger)}$ destroys (creates) an electron
with spin $\sigma$ at site
$i$, and $n_{i,\sigma}=c_{i,\sigma}^{\dagger}c_{i,\sigma}$. $U$ is the
on-site Hubbard repulsion.

Following KR \cite{KR} we enlarge the original Hilbert space by introducing
four subsidiary boson fields $e_{i}^{(\dagger)}$,
$s_{i,\uparrow}^{(\dagger)}$, $s_{i,\downarrow}^{(\dagger)}$,
and $d_{i}^{(\dagger)}$ for each site i.
These operators stand for the annihilation (creation) of
empty, singly occupied states with spin up or down, and doubly occupied
sites, respectively. Since there are only four possible states per site,
these boson projection operators must satisfy the completeness condition
\begin{equation}\label{CONST1}
e_{i}^{\dagger}e_{i}+\sum_{\sigma}s_{i,\sigma}^{\dagger}s_{i,\sigma}
+d_{i}^{\dagger}d_{i}=1
\end{equation}
Furthermore
\begin{equation}\label{CONST2}
n_{i,\sigma}=s_{i,\sigma}^{\dagger}s_{i,\sigma}+d_{i}^{\dagger}d_{i}
\end{equation}
Then, in the physical subspace defined by Eqs. (\ref{CONST1},\ref{CONST2})
the Hamiltonian (\ref{HM}) takes the form
\begin{equation}
\tilde{H}= \sum_{ij,\sigma}t_{ij}z_{i,\sigma}^{\dagger}f_{i,\sigma}^{\dagger}
f_{j,\sigma}z_{j,\sigma} + U\sum_{i}d_{i}^{\dagger}d_{i} \label{SB}
\end{equation}
with
\begin{equation}\label{eq:zfac}
z_{i,\sigma}=
e_{i}^{\dagger}s_{i,\sigma}+s_{i,-\sigma}^{\dagger}d_{i}
\end{equation}
and has the same matrix elements than those calculated for (\ref{HM}) in the
original Hilbert space. The operators $f_{i,\sigma}^{(\dagger)}$ are the
electron annihilation (creation) operators in the new Hilbert space. 

In the saddle-point approximation we can represent the wave-function
for a specific inhomogeneous solution $\alpha$ as
\begin{equation}\label{eq:ansatz}
|\Psi^\alpha\rangle = |\Phi_0^\alpha\rangle \otimes |B_0^\alpha\rangle
\end{equation}
where $|\Phi_0^\alpha\rangle$ is a Slater determinant and the 
bosonic part $|B_0^\alpha\rangle$ is a coherent state 
\begin{equation}
|B_0^\alpha\rangle= 
e^{\sum_i\left( \bar{d}_i^\alpha d_i^\dagger + \sum_\sigma \bar{s}_{i,\sigma}^\alpha 
s_{i,\sigma}^\dagger + \bar{e}_i^\alpha e_i^\dagger -1/2\right)}|0\rangle .
\end{equation}
Since a coherent state contains an arbitrary number of bosons the
constraints Eq. (\ref{CONST1},\ref{CONST2}) are only fulfilled
on average for a given inhomogeneous solution $\alpha$ provided that
\begin{eqnarray*}
1&=&(\bar{e}_{i}^{\alpha})^2+\sum_{\sigma}(\bar{s}_{i,\sigma}^{\alpha})^2
+(\bar{d}_{i}^{\alpha})^2 \\
\langle n_{i,\sigma}\rangle^\alpha &\equiv&  \langle \Phi_0^\alpha|n_{i,\sigma}|\Phi_0^\alpha\rangle 
 = (\bar{s}_{i,\sigma}^{\alpha})^2
+(\bar{d}_{i}^{\alpha})^2 .
\end{eqnarray*}
Note that here and in the following expectation
values of fermion operators are denoted with respect to the 
Slater determinant of $f$-electron operators.

The problem with the Ansatz Eq. (\ref{eq:ansatz}) is that one does not
recover the correct non-interacting limit $U\to 0$ for which $z_{i,\sigma}\to 1$. Therefore KR \cite{KR} introduced a unitary
transformation in order to represent the z-operators in Eq. (\ref{eq:zfac})
as
\begin{equation} \label{eq:znew}
z_{i,\sigma}=\frac{1}{\sqrt{e_{i}^{\dagger}e_{i}+s_{i,-\sigma}^{\dagger}
s_{i,-\sigma}}}(e_{i}^{\dagger}s_{i,\sigma}+s_{i,-\sigma}^{\dagger}d_{i})
\frac{1}{\sqrt{d_{i}^{\dagger}d_{i}+
s_{i,\sigma}^{\dagger}s_{i,\sigma}}}
\end{equation}
so that
\begin{equation}
\langle \Psi^\alpha|z_{i,\sigma}^{\dagger}c_{i,\sigma}^{\dagger}
c_{j,\sigma}z_{j,\sigma}|\Psi^\alpha\rangle
= (q^{\alpha}_i)^* q^\alpha_j \langle \Phi_0^\alpha|
c_{i,\sigma}^{\dagger}
c_{j,\sigma}|\Phi_0^\alpha\rangle .
\end{equation}
The expectation values of the z-operators Eq. (\ref{eq:znew}) 
\begin{equation}\label{krhop}
q^\alpha_{i,\sigma} = \langle B_0^\alpha| 
z_{i,\sigma}|B_0^\alpha\rangle 
\end{equation}
are equivalent to the renormalization factors derived within the
infinite ${\cal D}$ prescription 
of the Gutzwiller approximation \cite{GEBHARD} (cf. Eq. (\ref{eq:qfak})
in appendix \ref{secga}). 

In previous works \cite{sei98,sei982} we have proposed a method for
minimizing the KR
energy functional $E^\alpha=\langle \Psi^\alpha|H|\Psi^\alpha\rangle$
on finite clusters without imposing constraints with respect to translational and spin rotational
invariance. In the remainder of this section we evaluate the
matrix elements of the Hubbard model between two different inhomogeneous
solutions $|\Psi^\alpha\rangle$ which then will be used in order to
partially restore these symmetries.

We start with the overlap between wave-functions
belonging to different inhomogeneous solutions
\begin{eqnarray}\label{eq:sab2}
S_{\alpha\beta}&=&\langle \Psi^\alpha|\Psi^\beta\rangle  \\
&=&\langle\Phi_0^\alpha|\Phi_0^\beta\rangle \langle B_0^\alpha|B_0^\beta\rangle
\nonumber
\end{eqnarray}
where the overlap between coherent states reads as
\begin{equation}
\langle B_0^\alpha|B_0^\beta\rangle = 
e^{\sum_i\left( \bar{d}_i^\alpha \bar{d}_i^\beta + \sum_\sigma
  \bar{s}_{i,\sigma}^\alpha \bar{s}_{i,\sigma}^\beta 
 + \bar{e}_i^\alpha \bar{e}_i^\beta -1 \right)}.\label{ovbos}
\end{equation}

The fermionic overlap is given by
\begin{equation}
\langle\Phi_0^\alpha|\Phi_0^\beta\rangle
= \langle\Phi_0^\alpha|\Phi_0^\beta\rangle_\uparrow
\langle\Phi_0^\alpha|\Phi_0^\beta\rangle_\downarrow \\
\end{equation}
and the evaluation of the spin-dependent factors is outlined in appendix
\ref{secappb}.

We now proceed by calculating the matrix elements of the hamiltonian
Eq. (\ref{SB}) in the basis of the inhomogeneous wave-functions
$|\Psi^\alpha\rangle$. From the above definitions one obtains for the 
Hubbard interaction
\begin{equation}
\langle\Psi^\alpha|U \sum_{i}d_{i}^{\dagger}d_{i}|\Psi^\beta\rangle
= U \langle\Phi_0^\alpha|\Phi_0^\beta\rangle \langle B_0^\alpha|B_0^\beta\rangle
\sum_i \bar{d}_i^\alpha\bar{d}_i^\beta .
\end{equation}

The kinetic term is evaluated in a similar way as
\begin{equation}
\langle\Psi^\alpha|\hat{T}|\Psi^\beta\rangle
= \sum_{ij,\sigma}t_{ij} {z}_{i,\sigma}^{\alpha\beta}z_{j,\sigma}^{\beta\alpha}
\langle\Phi_0^\alpha|c_{i,\sigma}^{\dagger}
c_{j,\sigma}|\Phi_0^\beta\rangle \langle B_0^\alpha|B_0^\beta\rangle
\end{equation}
with the fermionic part
\begin{eqnarray}
\langle\Phi_0^\alpha|c_{i,\sigma}^{\dagger}
c_{j,\sigma}|\Phi_0^\beta\rangle
&=& \left\lbrack c_{i,\sigma}^{\dagger}
c_{j,\sigma}\right\rbrack_{\alpha\beta}
\langle\Phi_0^\alpha|\Phi_0^\beta\rangle_{-\sigma}
\end{eqnarray}
and the brackets are defined in Eq. (\ref{eq:det1}) in appendix \ref{secappb}.

The matrix elements of the 'bare' bosonic 'z'-operators 
from Eq. (\ref{eq:zfac}) read as
\begin{equation}
{z}_{i,\sigma}^{\alpha\beta}=
\bar{d}_i^\alpha\bar{s}_{i,-\sigma}^\beta 
+ \bar{s}_{i,\sigma}^\alpha\bar{e}_i^\beta . \label{eq:z1}
\end{equation}

Now we have to deal again with the problem that the z-factors as defined in
Eqs. (\ref{eq:z1}) do not yield the uncorrelated limit,
i.e. $\widetilde{z}_{i,\sigma}^{\alpha\beta}=z_{i,\sigma}^{\alpha\beta}\to 1$
for $U\to 0$. It is straightforward to proof that the representation of
Eq. (\ref{eq:znew}) does not work in this case since the above limit is
only obeyed for homogeneous paramagnetic solutions.
However, due to a non-symmetric population of momentum states 
on finite clusters or in case of inclusion of an electron-phonon coupling 
the charge and spin structure in general is inhomogeneous even in the
limit $U\to 0$.
 
A possible representation which yields $z_{i,\sigma}^{\alpha\beta}\to 1$
for $U\to 0$ is given by
\begin{widetext}
\begin{eqnarray}\label{eq:hoprenorm}
z_{i,\sigma}^\dagger&=&\frac{1}{\sqrt{1-e_{i}^{\dagger}e_{i}^\dagger
-s_{i,-\sigma}^{\dagger}s_{i,-\sigma}^\dagger}}\left\lbrack
\sqrt{1-e_{i}^{\dagger}e_{i}^\dagger
-s_{i,\sigma}^{\dagger}s_{i,\sigma}^\dagger}\,\,
d_{i}^{\dagger}s_{i,-\sigma}\,\,
\frac{1}{\sqrt{1-e_{i}e_{i}-s_{i,\sigma}s_{i,\sigma}}} \right. \nonumber \\
 &+& \left.\sqrt{1-d_{i}^{\dagger}d_{i}^\dagger
-s_{i,-\sigma}^{\dagger}s_{i,-\sigma}^\dagger}\,\,
s_{i,\sigma}^{\dagger}d_{i}\,\,
\frac{1}{\sqrt{1-d_{i}d_{i}-s_{i,-\sigma}s_{i,-\sigma}}}
\right\rbrack
\frac{1}{\sqrt{1-d_{i} d_{i}-
s_{i,\sigma} s_{i,\sigma}}} .
\end{eqnarray}
\end{widetext}
Note that in the physical subspace defined by Eq. (\ref{CONST1})
the square root factors are identically 'one'.
On the other hand, upon evaluating 
the matrix elements of Eq. (\ref{eq:hoprenorm}) between coherent states
$\alpha,\beta$ one obtains  the hopping renormalization factors
\begin{eqnarray}
&{z}_{i,\sigma}^{\alpha\beta}&\equiv 
\langle B_0^\alpha|z_{i,\sigma}^{\dagger}|B_0^\beta\rangle 
=\frac{1}
{\sqrt{\langle n_{i,\sigma}\rangle_\alpha(1-\langle 
n_{i,\sigma}\rangle_\beta)}}\!\!\! \label{zrenorm} \\
&\times&\!\!\! \left\lbrace \sqrt{\frac{\langle n_{i,-\sigma}\rangle_\alpha}{\langle n_{i,-\sigma}\rangle_\beta}}
\sqrt{(\bar{d}^\alpha_i)^2(\langle n\rangle_{i,-\sigma}^\beta 
-(\bar{d}_i^\beta)^2)} \right. \nonumber \\
&+&\!\!\!\left. \sqrt{\frac{1-\langle n_{i,-\sigma}\rangle_\alpha}{1-\langle n_{i,-\sigma}\rangle_\beta}}
\sqrt{(1-\langle n\rangle_{i}^\beta +(\bar{d}_i^\beta)^2)(\langle
  n\rangle_{i,\sigma}^\alpha -(\bar{d}_i^\alpha)^2)}\right\rbrace \nonumber 
\end{eqnarray}
where we have used the constraints Eqs. (\ref{CONST1},\ref{CONST2}) to replace
the boson fields but $\bar{d}_i^{\alpha}$ by fermionic 
expectation values. The 'z-factors' Eq. (\ref{zrenorm}) show
the correct behavior ${z}_{i,\sigma}^{\alpha\beta}\to 1$ for
$U\to 0$ and the diagonal elements 
reduce to the KR renormalization factors Eq. (\ref{krhop}), i.e. 
${z}_{i,\sigma}^{\alpha\alpha}=q^\alpha_{i,\sigma}$. 

In appendix \ref{secga} it is shown that the renormalization factors
Eq. (\ref{zrenorm}) can be also motivated from the generalized 
Gutzwiller approach in the limit ${\cal D}\to \infty$.

\section{Results}
In the previous section we have calculated the
matrix elements between different inhomogeneous
states $|\Psi^\alpha\rangle$ of the Hubbard model.
These results are now used for
evaluating an improved ground state energy and wave-function
similar than in the configuration interaction approach
based on unrestricted HF wave-functions.~\cite{louis01}

We apply the method to the investigation of spin polaron states on
a square lattice,
i.e. we have one hole with respect to half-filling.
Minimization of the KR (or GA) energy functional 
leads to the localization of this hole at a given site  $R_\alpha$
(cf. Ref. \onlinecite{sei98} for a method of performing the
unrestricted variation) 
and we denote the corresponding projected or fermion-boson wave-function with  
$|\Psi^\alpha\rangle$.

Now we generate all translations  
of this solution within the same sublattice since solutions belonging 
to different sublattices are orthogonal. 
The superposition
\begin{equation}\label{eq:sup}
|\Psi\rangle = \sum_\alpha v_\alpha |\Psi^\alpha\rangle
\end{equation}
thus only includes states $|\Psi^\alpha\rangle$ with the same
energy $E=E^\alpha$.
In principle one could systematically improve
the approach by including also excited states of the underlying
fermionic Slater determinant. 

If we apply the hamiltonian Eq. (\ref{HM}) to Eq. (\ref {eq:sup})
one obtains the following eigenvalue problem
\begin{equation}\label{eq:math}
\langle \Psi^\alpha|H|\Psi^\beta\rangle v_\beta = \varepsilon
S_{\alpha\beta}v_\beta
\end{equation}
where the matrix $S_{\alpha\beta}$ is defined in 
Eq. (\ref{eq:sab2}).

\subsection{One hole states in the 4x4 lattice}
We start by investigating the quality of the present approach
with regard to exact results and the HF configuration 
interaction method (CIHF).

Table \ref{tab1} reports the energy correction obtained with
our slave-boson configuration interaction approach (CISB) 
as compared to the unrestricted GA. The values for the exact result, the
CIHF and the unrestricted HF (from 
Ref. \onlinecite{louis99}) are also shown for comparison.

\begin{table}[htp] \label{tab1}
\begin{tabular}{|l|l|l|l|l|l|}\hline
U/t & exact & HF & GA & CIHF & CISB \\ \hline \hline
4 & -0.91658 & -0.83139 & -0.88815 & -0.83501 & -0.89091 \\ \hline 
6 & -0.74794 & -0.64222 & -0.70020 & -0.66214 & -0.70497 \\ \hline
8 & -0.634203 & -0.52884 & -0.57518 & -0.54767 & -0.60295 \\ \hline
16 & -0.42546 & -0.33589 & -0.37130 & -0.34604 & -0.38091 \\ \hline
32 & -0.308473 & -0.23160 & -0.27209 & -0.23627 & -0.27685 \\ \hline
50 & -0.266039 & -0.19335 & -0.23954 & -0.19617 &  -0.24362 \\ \hline
\end{tabular}
\caption{Energy per site for 15 particles on a $4\times 4$ lattice. The values of
the exact result, HF and CIHF method have been taken from 
Ref. \onlinecite{louis99}}
\end{table}

It turns out that the CISB leads to an energy correction to the GA
result which is of the same order of magnitude than the CIHF
correction to the HF energy. However, this improvement is on
top of the  GA which itself provides a much
better estimate for the ground state energy than the HF approximation.
For example, one finds that for $U/t=8$ the CISB differs from
the exact result by $\approx 5\%$ whereas it is $\approx 13\%$ in case
of the CIHF.

\subsection{One hole states in the 16x16 lattice}

We continue by evaluating the dispersion of the spin polaron 
in a $16\times 16$ lattice. This problem has been extensively
investigated within the tJ model, 
\cite{horsch,tohyama,vojta,liu,reiter,leung,rink,marsiglio,kane}
where for small $J/t$
one finds a bandwidth $\sim J$ which turns over into a
$2t^2/J^4$ behavior for large $J/t$. Furtheron the dispersion is
characterized by a maximum at $(0,0)$ (and the analogous $(\pi,\pi)$
point) and displays a 'hole pocket' at $(\pi/2,\pi/2)$ which
is slightly lower in energy than the $(\pi,0)$ point. 

\begin{figure}[tbp]
\includegraphics[width=8cm,clip=true]{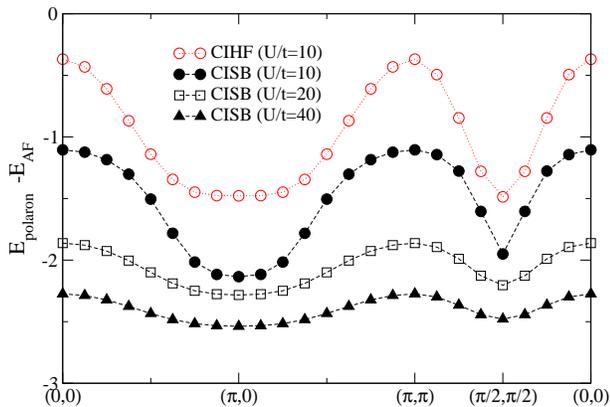}
\caption{(color online). Dispersion of the spin polaron in the Hubbard model evaluated
within the CISB ($U/t=10,20,40$) and CIHF ($U/t=10$) method. Energies are
with respect to the half-filled antiferromagnet.}
\label{fig2}
\end{figure}

Fig. \ref{fig2} displays the polaron dispersion obtained within the
SBCI method for $U/t=10,20,40$. For comparison we also show the
$U/t=10$ result obtained from the CIHF method.
Since the wave-function incorporates only polaron states
localized on the same sublattice the dominant contribution to 
the dispersion is given by
$E_k \approx 4t'\cos(k_x)\cos(k_y) + 2t[\cos(2 k_x)+\cos(2 k_y)]$.
Therefore at the point $k=(\pi,\pi/2)$ the energy difference between
CIHF and CISB corresponds to the difference between GA and HF energies
for the spin polaron.
Since within the CISB approach the matrix elements which enter
Eq. (\ref{eq:math}) are additionally scaled by the bosonic exponential
overlap Eq. (\ref{ovbos}) the corresponding long range contributions to the 
dispersion are in generally smaller than for the CIHF method. 
On the other hand, this scaling affects also the matrix 
$S_{\alpha\beta}$ in Eq. (\ref{eq:math}) so that due to
partial cancellation the overall
effect on the bandwidth is less pronounced as one might expect (see below).

From analogous investigations in the tJ-model 
\cite{horsch,tohyama,vojta,liu,reiter,leung,rink,marsiglio,kane} it
is known that the dispersion of a single hole has a saddle-point at
$k=(\pi,0)$  and  $k=(\pi/2,\pi/2)$, where the latter corresponds to
the minimum of the band.
From Fig. \ref{fig2} it turns out that the CIHF spin-polaron dispersion  
also displays the minimum at $k=(\pm \pi/2,\pm \pi/2)$ whereas within the
CISB method the state at $k=(\pm \pi,0), (0,\pm \pi)$ 
is slightly lower in energy.
However, a direct comparison of results between tJ- and Hubbard model is 
hampered by the fact that
the strong coupling expansion of the Hubbard model generates
a three-site term of order $J$ in addition to the 'conventional' tJ-model.
Since we find that the energy difference between 
$k=(\pm \pi/2,\pm \pi/2)$ and $k=(\pm \pi,0), (0,\pm \pi)$ states
is always smaller than $J=4t^2/U$
there appears no inconsistency with results from the tJ-model. 
In fact, calculations of a single hole in the antiferromagnet based on 
an expanded tJ-model (including the three-site term) 
provide evidence that the minimum of the band may be
at $k=(\pm \pi,0), (0,\pm \pi)$ \cite{elrick}. 
This finding is also substantiated by
exact diagonalization results of the same model on small clusters 
.~\cite{fehske} Unfortunately, for the full Hubbard model
there are no conclusive answers from Quantum Monte Carlo  
or exact methods yet available.~\cite{sorella,dagotto}

\begin{table}[htp] 
\begin{tabular}{|l|l|l|l|l|l|l|}\hline
& \multicolumn{2}{c|}{\it SCBA} & \multicolumn{2}{c|}{\it CIHF} &
\multicolumn{2}{c|}{\it CISB} \\ \hline  
J & $E_{(\pi/2,\pi/2)}$ & W & $E_{(\pi/2,\pi/2)}$ & W & 
$E_{(\pi/2,\pi/2)}$ &W \\ \hline \hline
0.1 & -2.785 & 0.239 & -1.84 & 0.231 &  -2.4786 & 0.263 \\ \hline
0.2 & -2.540 & 0.430 & -1.703 & 0.513 & -2.204 & 0.421\\ \hline
0.3 & -2.360 & 0.600 & -1.588 & 0.817 & -2.036 & 0.68\\ \hline
0.4 & -2.209 & 0.741 & -1.487 & 1.118 &  -1.95 & 1.031 \\ \hline
\end{tabular}
\caption{Binding energy $E_{polaron}-E_{AF}$ taken at
momentum $q=(\pi/2,\pi/2)$ and the bandwidth $W$ for various values
of $J=4t^2/U$. Shown are results for the self-consistent Born approximation
(SCBA) of the tJ-model (from Ref. \onlinecite{horsch}) and the CIHF and CISB 
method for the Hubbard model, respectively.} \label{tab2}
\end{table}

Table \ref{tab2} reports the bandwidth, and the energy at $k=(\pm \pi/2,\pm
\pi/2)$ of the spin polaron dispersion obtained within the SCBA,~\cite{horsch}
CIHF and CISB method, respectively. Note that for the latter
approach the bandwidth is $W=E_{(0,0)}-E_{(\pi,0)}$ whereas for the
SCBA and CIHF methods it is given by $W=E_{(0,0)}-E_{(\pi/2,0)}$.
Despite this difference we find that the CISB bandwidth scales
as $W \approx 2.2 J $ up to $J \approx 0.3$ in agreement with 
analogous considerations in the tJ-model.
It also turns out that (at least for $J>0.1$) the CISB bandwidth
is smaller than that of the CIHF approach. Formally this is again due
to the additional renormalization of the matrix elements 
by the bosonic exponential overlap Eq. (\ref{ovbos}). On the other
hand it is quite natural that the CISB approach leads to 
'heavier' spin polarons than the CIHF method due to the 
incorporation of correlation effects already on the Gutzwiller level. 
Similar to the case of the $4\times 4$ lattice the CISB leads to a significant
energy correction with regard to the CIHF as exemplified by the value of
$E_{(\pi/2,\pi/2)}$ in table \ref{tab2}.

\subsection{Comparison with experiment}
\begin{figure}[tbp]
\includegraphics[width=8cm,clip=true]{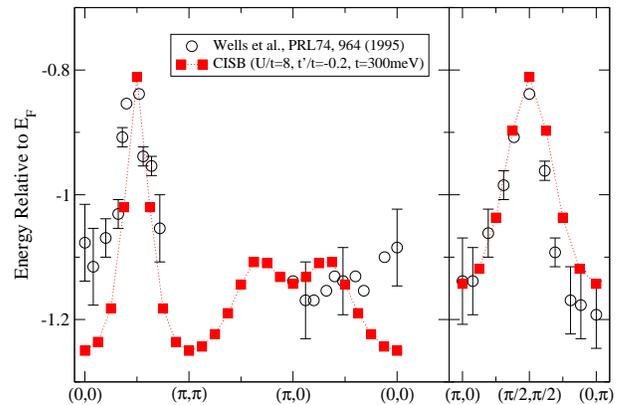}
\caption{(color online). Dispersion of the spin polaron in the extended Hubbard model evaluated
within the CISB ($U/t=8$, $t'/t=-0.2$, $t=300 meV$). 
The right panel shows the direction along the boundary of the magnetic
Brillouin zone. Experimental data are from Ref. \onlinecite{wells}.}
\label{fig3}
\end{figure}

Undoped cuprate superconductors are antiferromagnetic Mott insulators.
Within a angle-resolved photoemission (ARPES) experiment, 
one can in principle observe
the dispersion of the created hole in the antiferromagnetic
background of these compounds and compare with that of of the spin polaron 
quasiparticle concept from the previous section.
On the basis of the single-band description it is now well established
from LDA \cite{Pavarini} and the analysis of ARPES data \cite{tanaka} 
that a next-nearest neighbor hopping $t'$ has to be considered in the
model. In particular, it has been found \cite{tanaka} that the 
quasiparticle dispersion 
from  $(\pi,0)$ to $(\pi/2,\pi/2)$, which is determined by $t'$, 
is characteristic for the different cuprate families. 
Our analysis below is therefore based on the extended Hubbard model,
which corresponds to Eq. (\ref{HM}) when the hopping $t_{ij}$
is restricted to nearest $\sim t$ and next-nearest $\sim t'$ hopping.
In Fig. \ref{fig3} we fit the resulting spin polaron 
dispersion to ARPES data on undoped
Sr$_2$CuO$_2$Cl$_2$ obtained  Wells {\it et al.}.~ \cite{wells}
Since the experiment measures the single particle Green's function
for electrons the dispersion in Fig. \ref{fig3} is 'reversed' with respect
to those shown in Fig. \ref{fig2} which were obtained for holes.

We can use the experimental energy differences 
$\Delta E_1 = E_{(\pi/2,\pi/2)} - E_{(0,0)}$ and 
$\Delta E_2=E_{(\pi/2,\pi/2)}-E_{(\pi,0)}$
in order to fit two of the three parameters ($t$, $t'$, $U$).
Therefore we additionally use our results from 
Ref. \onlinecite{goetz05} where we have fitted the magnon dispersion of
undoped La$_2$CuO$_4$ within the time-dependent Gutzwiller approximation. 
In this case the value of the Hubbard repulsion $U/t\approx 8$ 
could be accurately
determined from the dispersion of spin excitations along the magnetic
Brillouin zone whereas this dispersion is rather unsensitive to $t'$.
Given that the Cu onsite repulsion should not depend very much on
the material we there also use the ratio $U/t$ in our present fit
of the spin polaron dispersion for Sr$_2$CuO$_2$Cl$_2$.
As a result we find that the ratio $t'/t=-0.2$ 
yields an overall good
agreement with the data and the nearest neighbor hopping $t=300meV$
is set by the absolute energy scale.
The ARPES data in addition allow
for an accurate determination of $t'$ so that a combination
of both approaches in principle can be used to obtain parameter
sets for the Hubbard model in order to describe different materials.

\section{Conclusions}
We have developed a configuration interaction approach based
on the KR slave-boson mean-field formulation
of the Hubbard model \cite{KR}. In principle this method provides a controlled
scheme for including fluctuations beyond the mean-field solution.
Formally this has been achieved by several authors within the functional
integral formalism.~\cite{ARRIGONI,LAVAGNA,LI,rai93,rai95}
Here we have discussed an alternative extension which is based on the
observation that unrestricted variation of the
KR energy functional in general leads to a class of degenerate solutions which
are connected by symmetry transformations. The CISB method discussed
in this paper allows for a tunneling between these degenerate solutions
and thus for a construction of eigenstates with well defined momentum.

Although the KR mean-field energy functional is identical to the
that obtained with the generalized Gutzwiller wave-function in 
${\cal D}\to\infty$ \cite{GEBHARD} the considerations in
appendix \ref{secga} show that the latter
approach leads to an 'orthogonality catastrophe' for matrix
elements between {\it different} inhomogeneous states. 
Therefore one would have to invoke $1/{\cal D}$ corrections 
in order to construct a CI approach
also within the Gutzwiller method. 

Application of the CISB to the spin-polaron problem for the Hubbard model 
leads to a significant
energy gain with respect to the CIHF method. In addition we have
obtained a minimum of the spin polaron dispersion at $k=(\pm \pi,0),
(0,\pm \pi)$ in contrast to analogous calculations in the tJ-model
but also in contrast to the CIHF method. However, calculations
based on the full strong coupling expansion of the Hubbard model 
,~\cite{elrick,fehske} which
take into account the three-site terms of order $t^2/U$, 
neglected in the 'conventional' tJ-model,  indicate the occurence 
of dispersion minima around the corners of the magnetic Brillouin zone.
To our knowledge there are no recent exact diagonalization studies 
of one hole in a $\sqrt{18}\times \sqrt{18}$ or $\sqrt{20}\times \sqrt{20}$ 
Hubbard cluster which could substantiate the findings of
Ref. \onlinecite{fehske}. However, since on the mean-field level 
the KR slave-boson formulation of the Hubbard model takes 
into account correlations beyond HF  we expect that the CISB is more 
accurate concerning fine details of the spin polaron dispersion as compared
to the CIHF method. Further investigations are needed in order
to confirm the finding of one hole dispersion minima at $k=(\pm \pi,0),
(0,\pm \pi)$ in the Hubbard model.  
 
Finally, we have included a next-nearest neighbor hopping $t'/t<0$ 
in the bare hamiltonian in order to fit the low enery dispersion of
Sr$_2$CuO$_2$Cl$_2$ from ARPES experiments.~\cite{wells}
The parameter $t'$ is essential in order to obtain the measured
dispersion along the border of the magnetic Brillouin zone.
More recent ARPES experiments \cite{duerr} have also revealed a 
strong dispersion along the $(0,0)\to (\pi,0)$ direction.
Within a one-band description modeling of these data requires
inclusion of a significant third nearest neighbor hopping. However, since
our CISB approach can be implemented also on the more realistic
three-band model it would be interesting to study the spin polaron 
dispersion within this hamiltonian. The comparison with ARPES experiments
would then allow to elucidate the parameters of this hamiltonian for
different cuprate materials. Moreover, since the superposition
in Eq. (\ref{eq:sup}) can be extended to  include also excited states, it should
be possible to calculate also the incoherent part of the ARPES spectrum
and thus to provide a more detailed description of the data.
Work in this direction is in progress.

\acknowledgments
I'm indepted to J. Lorenzana for a critical reading of the manuscript
any many valuable comments. I also thank V. Hizhnyakov 
for helpful discussions.

\appendix
\section{Generalized Gutzwiller approximation}\label{secga}
Following Ref. \onlinecite{GEBHARD} the Ansatz for a given inhomogeneous 
state $\alpha$ can be written as
\begin{eqnarray}\label{psivar} 
|\Psi^\alpha\rangle &=& g^{\hat{K(\alpha)}}|\Phi^\alpha_0\rangle
=\prod_i\hat{B}_i^\alpha|\Phi_0^\alpha\rangle \\
\hat{B}_i^\alpha &=& g^{\hat{K}_i(\alpha)} = g^{\hat{D}_i-\sum_\sigma \mu^\alpha_{i,\sigma}
\hat{n}_{i,\sigma} + \eta_i^\alpha}\label{eq:bi}
\end{eqnarray}
where the uncorrelated state $|\Phi_0^\alpha\rangle$ is a Slater-determinant
with an inhomogeneous density matrix $\alpha$
and $\hat{D_i}=n_{i,\uparrow}n_{i,\downarrow}$
is the double occupancy operator.
For later purposes we also define the operators for single occupied (with
spin $\sigma$) and empty sites:
\begin{eqnarray}
\hat{S}_{i,\sigma} &=& \hat{n}_{i,\sigma}(1-\hat{n}_{i,-\sigma}) \\
\hat{E}_i &=& (1-\hat{n}_{i,\sigma})(1-\hat{n}_{i,-\sigma}).
\end{eqnarray}
The parameters 
$\mu^\alpha_{i,\sigma}$ and $\eta_i^\alpha$ have to be determined 
variationally.
Gebhard \cite{GEBHARD} has shown that the requirement
\begin{equation}\label{eq:iden}
g^{2\hat{K}(\alpha)} \equiv \sum_i ln \left\lbrack 1+x^\alpha_i
(\hat{D}_i-{D}_i^{HF,\alpha})\right\rbrack
\end{equation}
leads to the same energy functional than the Kotliar-Ruckenstein slave-boson
approach in the mean-field approximation when the expectation values
are formally evaluated in the limit of infinite dimensions.
Here ${D}_i^{HF,\alpha}$ denotes the Hartree-Fock decoupled double
occupancy operator in the basis of the Slater determinant 
$|\Phi_0^\alpha\rangle$.
Eq. (\ref{eq:iden}) yields a relation between the variational parameters
$g$, $\mu^\alpha_{i,\sigma}$, $\eta_i^\alpha$ and the variables
$x^\alpha_i$ which turn out to be the relevant parameters when one
evaluates expectation values in infinite dimensions.
The essential step in this direction is to express the operator 
$\hat{B}_i^\alpha$ defined in Eq. 
(\ref{eq:bi}) in terms of the $x^\alpha_i$ as
\begin{eqnarray}
\hat{B}_i^\alpha &=& \hat{D}_i\sqrt{1+x_i^\alpha \langle{E}_i\rangle^{\alpha}} 
+\sum_\sigma \hat{S}_{i,\sigma}
\sqrt{1-x_i^\alpha \langle{S}_{i,-\sigma}\rangle^{\alpha}}\nonumber \\  
&+& \hat{E}_i\sqrt{1+x_i^\alpha \langle{D}_i\rangle^{\alpha}}. \label{eq:bi2}
\end{eqnarray}
and the expectation values are defined with regard to $|\Phi_0^\alpha\rangle$.
An important result of the $d\to \infty$ description is the equivalence
of local densities in the projected and unprojected states
\begin{equation}\label{eq:equiv}
\langle \Psi^\alpha|c_{i,\sigma}^\dagger c_{i,\sigma}|\Psi^\alpha\rangle
= \langle \Phi_0^\alpha|c_{i,\sigma}^\dagger c_{i,\sigma}|\Phi_0^\alpha\rangle
\end{equation}
which will be used in the following.

First the double occupancy can be evaluated as
\begin{equation}\label{eq:dble}
\langle \Psi^\alpha|\hat{D_i}|\Psi^\alpha\rangle \equiv {\cal D}^\alpha_i
= \langle{D}_i\rangle^{\alpha}(1+x_i^\alpha \langle{E}_i\rangle^{\alpha})
\end{equation}
which allows one to perform the variations with respect to the double
occupancy ${\cal D}^\alpha_i$ instead of $x_i^\alpha$ 
(or $g$, $\mu^\alpha_{i,\sigma}$, and $\eta_i^\alpha$).
Analogously the hopping term of Eq. (\ref{HM}) is given by
\begin{equation}
\langle \Psi^\alpha|c_{i,\sigma}^\dagger c_{j,\sigma}|\Psi^\alpha\rangle
= q_{i,\sigma}^\alpha q_{j,\sigma}^\alpha\langle \Phi^\alpha_0|
c_{i,\sigma}^\dagger c_{j,\sigma}|\Phi^\alpha_0\rangle
\end{equation}
with the hopping renormalization factors
\begin{eqnarray}
q^\alpha_{i,\sigma} &=&  \frac{1-\langle n_{i,-\sigma}\rangle^\alpha}
{\sqrt{\langle{E}_{i}\rangle^{\alpha}\langle{S}_{i,\sigma}\rangle^{\alpha}}}
\sqrt{{\cal S}_{i,\sigma}^\alpha {\cal E}_i^\alpha} \nonumber \\
&+&\frac{\langle n_{i,-\sigma}\rangle^\alpha}
{\sqrt{\langle{D}_{i}\rangle^{\alpha}\langle{S}_{i,-\sigma}\rangle^{\alpha}}}
\sqrt{{\cal D}_i^\alpha {\cal S}_{i,-\sigma}^\alpha }
\label{eq:qfak}
\end{eqnarray}
Similar than in Eq. (\ref{eq:dble}) expectation values  
of a projection operator $\hat{P}_i=\hat{D}_i$, $\hat{S}_{i,\sigma}$, 
$\hat{E}_i$ with regard to $|\Psi^\alpha\rangle$ have been
denoted with calligraphic letters.  
The above representation of the hopping factors allows 
for a interpretation of the renormalized kinetic energy in terms
of 'probability ratios'.~\cite{VOLLHARDT,zhang88}
Consider the term $q^\alpha_{i,\sigma} c_{i\sigma} |\Phi_0^\alpha\rangle$
which is the sum of two processes: 
The contribution $\sim c_{i\sigma} (1-\langle
n_{i,-\sigma}\rangle^\alpha)$ originates from the annihilation 
of a singly occupied
(and thus creation of an empty) site and is weighted by the ratios 
between projected and unprojected probabilities of this process.
The contribution $\sim c_{i\sigma} \langle n_{i,-\sigma}\rangle^\alpha$
weights in a similar way the annihilation of an electron on a 
doubly occupied site.

We now proceed by evaluating the matrix ${\bf S}$ which contains the overlap 
elements of wave-functions belonging to different inhomogeneous states
\begin{eqnarray}\label{eq:sab}
S_{\alpha\beta}&=&\langle \Psi^\alpha|\Psi^\beta\rangle
= \prod_i \langle \Phi_0^\alpha|\hat {B}_i^\alpha
\hat {B}_i^\beta|\Phi_0^\beta\rangle \\
&=& \prod_i\left\lbrace \sum_\sigma\sqrt{\frac{{\cal S}_{i,\sigma}^{\alpha}
{\cal S}_{i,\sigma}^{\beta}}
{\langle{S}_{i,\sigma}\rangle^{\alpha}
\langle{S}_{i,\sigma}\rangle^{\beta}}}
\langle 
\Phi_0^\alpha|\hat {S}_{i,\sigma}|\Phi_0^\beta\rangle \right. \nonumber \\
&+& \sqrt{\frac{{\cal E}_{i}^{\alpha}
{\cal E}_{i}^{\beta}}
{\langle{E}_{i}\rangle^{\alpha}
\langle{E}_{i}\rangle^{\beta}}}
\langle 
\Phi_0^\alpha|\hat{E}_{i}|\Phi_0^\beta\rangle  \nonumber \\
&+& \left.\sqrt{\frac{{\cal D}_{i}^{\alpha}
{\cal D}_{i}^{\beta}}
{\langle{D}_{i}\rangle^{\alpha}
\langle{D}_{i}\rangle^{\beta}}}
\langle 
\Phi_0^\alpha|\hat {D}_{i}|\Phi_0^\beta\rangle \right\rbrace \nonumber
\end{eqnarray}
where we have  used Eqs. (\ref{eq:bi2},\ref{eq:equiv},\ref{eq:dble})
and the fact that only local contractions survive in infinite dimensions.
Eq. (\ref{eq:sab}) also requires the evaluation of matrix
elements of $\hat{P}_i$  between different Slater determinants 
$\langle \Phi_0^\alpha|\hat {P}_i|\Phi_0^\beta\rangle$.  
For example, one finds for the double occupancy
operator
\begin{equation}\label{eq:dble2}
\langle \Phi_0^\alpha|\hat {D}_i|\Phi_0^\beta\rangle
= \left\lbrack \hat{n}_{i,\uparrow}\right\rbrack_{\alpha\beta}
\left\lbrack \hat{n}_{i,\downarrow}\right\rbrack_{\alpha\beta}
\end{equation}
and the brackets are defined in Eq. (\ref{eq:det1}).

Schwartz's inequality together with the relation between harmonic and 
geometric mean
\begin{eqnarray}
\langle\Phi_0^\alpha|\hat{P}_{i}|\Phi_0^\beta\rangle &\le& 
\sqrt{\langle{P}_{i}\rangle^{\alpha}
\langle{P}_{i}\rangle^{\beta}} \\
\sqrt{{\cal P}_i^\alpha{\cal P}_i^\beta} &\le& 
({\cal P}_i^\alpha + {\cal P}_i^\beta)/2 
\end{eqnarray} 
yields
\begin{equation}
\langle \Phi_0^\alpha|\hat {B}_i^\alpha
\hat {B}_i^\beta|\Phi_0^\beta\rangle \le 1
\end{equation}
where the equals sign holds for $\alpha=\beta$.


Analogously to ${\bf S}$ one can evaluate the matrix elements of the
Hubbard hamiltonian Eq. (\ref{HM}).
For the double occupancy operator one obtains
\begin{eqnarray}\label{eq:doub}
\langle \Psi^\alpha|\hat{D}_i|\Psi^\beta\rangle
&=& \frac{\langle \Phi_0^\alpha|\hat{B}_i^\alpha\hat{D}_{i}
\hat{B}_i^\beta|\Phi_0^\beta\rangle}
{\langle \Phi_0^\alpha|\hat{B}_i^\alpha
\hat{B}_i^\beta|\Phi_0^\beta\rangle} S_{\alpha\beta} \\
&=& \sqrt{\frac{{\cal D}_i^\alpha{\cal D}_i^\beta}
{\langle{D}_{i}\rangle^{\alpha}\langle{D}_{i}\rangle^{\beta}}}
\frac{\langle \Phi_0^\alpha|\hat{D}_{i}|\Phi_0^\beta\rangle}
{\langle \Phi_0^\alpha|\hat{B}_i^\alpha
\hat{B}_i^\beta|\Phi_0^\beta\rangle} S_{\alpha\beta}\nonumber 
\end{eqnarray}
and the matrix elements of the hopping term are given by
\begin{equation}\label{eq:hop}
\langle \Psi^\alpha|c^\dagger_{i,\sigma}c_{j,\sigma}|\Psi^\beta\rangle
= \frac{\langle \Phi_0^\alpha|
\hat{B}^\alpha_{i}c^\dagger_{i,\sigma}\hat{B}^\beta_{i}
\hat{B}^\alpha_{j}c_{j,\sigma}\hat{B}^\beta_{j}
|\Phi_0^\beta\rangle}
{\langle \Phi_0^\alpha|\hat{B}_i^\alpha
\hat{B}_i^\beta|\Phi_0^\beta\rangle
\langle \Phi_0^\alpha|\hat{B}_j^\alpha
\hat{B}_j^\beta|\Phi_0^\beta\rangle} S_{\alpha\beta}.
\end{equation}

Using Eqs. (\ref{eq:bi2},\ref{eq:equiv},\ref{eq:dble}) the projections of 
the creation and annihilation operators can be expressed as
\begin{eqnarray}
\hat{B}^\alpha_{i}c^\dagger_{i,\sigma}\hat{B}^\beta_{i}
&=& \left\lbrack (1-n_{i,-\sigma})\sqrt{\frac{{\cal S}_{i\sigma}^\alpha
{\cal E}_i^\beta}{\langle S_{i\sigma}\rangle^\alpha\langle E_{i}\rangle^\beta}}
\right. \nonumber \\
&+& \left.n_{i,-\sigma}  \sqrt{\frac{{\cal D}_{i}^\alpha
{\cal S}_{i,-\sigma}^\beta}{\langle D_{i}\rangle^\alpha\langle S_{i,-\sigma}
\rangle^\beta}}\right\rbrack c_{i,\sigma}^\dagger  
\label{eq:pro1}\\
\hat{B}^\alpha_{j}c_{j,\sigma}\hat{B}^\beta_{j}
&=& \left\lbrack (1-n_{j,-\sigma})\sqrt{\frac{{\cal E}_{j}^\alpha
{\cal S}_{j\sigma}^\beta}{\langle E_{j}\rangle^\alpha\langle S_{j\sigma}
\rangle^\beta}}
\right. \nonumber \\
&+& \left.n_{j,-\sigma}  \sqrt{\frac{{\cal S}_{j,-\sigma}^\alpha
{\cal D}_{j}^\beta}{\langle S_{j,-\sigma}\rangle^\alpha\langle D_{j}
\rangle^\beta}}\right\rbrack c_{j,\sigma}\label{eq:pro2}
\end{eqnarray}

In principle it is possible to evaluate the matrix elements from
Eqs. (\ref{eq:hop}) in terms of the Slater determinants
$|\Phi_0^\alpha\rangle$, however, the calculation of contributions 
which involve
density correlations of the form $\langle \Phi_0^\alpha|n_{i,-\sigma}
n_{j,-\sigma}|\Phi_0^\beta\rangle$ are rather time consuming.
We therefore simplify the expression of
the projections Eqs. (\ref{eq:pro1},\ref{eq:pro2}) by the following argument. 
With regard to the matrix element Eq. (\ref{eq:hop}) 
the projection Eq. (\ref{eq:pro1}) describes the annihilation of a 
particle with spin $\sigma$
in the Slater determinant $\langle \Phi_0^\alpha|$. The two contributions
measure the probabilty wether site $i$ in the state $\alpha$ is singly or
doubly occupied. Accordingly we replace the corresponding projections
by their mean-values, e.g.  $1-n_{i,-\sigma} \to 1-\langle
n_{i,-\sigma}\rangle^\alpha$. In the same way Eq. (\ref{eq:pro2}) describes
the annihilation of a particle with spin $\sigma$ in the Slater-determinant
$|\Phi_0^\beta\rangle^\beta$ and we approximate in this case 
$1-n_{i,-\sigma} \to 1-\langle
n_{i,-\sigma}\rangle^\beta$.
Within this approximation one obtains for the projected creation and 
annihilation operators
\begin{eqnarray}
\hat{B}^\alpha_{i}c^\dagger_{i,\sigma}\hat{B}^\beta_{i}
&=& {q}^{\alpha\beta}_{i,\sigma} c_{i,\sigma}^\dagger \\     
\hat{B}^\alpha_{j}c_{j,\sigma}\hat{B}^\beta_{j}
&=& {q}^{\beta\alpha}_{j,\sigma} c_{j,\sigma}
\end{eqnarray}
where the ${q}^{\alpha\beta}_{i,\sigma}$ are equivalent to the renormalization factors
Eq. (\ref{zrenorm}) derived with the KR slave-boson method.

In case of the GA we observe from Eq. (\ref{eq:sab}) that  
$S_{\alpha\ne \beta}$ is a product over lattice sites of 
terms less than 'one' which in the thermodynamic limit 
leads to an 'orthogonality
catastrophe' \cite{and67} and thus $S_{\alpha\beta}=\delta_{\alpha\beta}$.
Therefore we find that within the 'infinite D' prescription of the
Gutzwiller approximation \cite{GEBHARD} different inhomogeneous
states are orthogonal to each other.
As a consequence it turns out from Eqs. (\ref{eq:doub},\ref{eq:hop}) that
these states are not connected by matrix elements of the Hubbard hamiltonian
so that a CI approach does not yields any correction to
the symmetry-broken solutions.

\section{Fermionic matrix elements}\label{secappb}
When we restrict to collinear inhomogeneous Gutzwiller solutions, i.e. 
where the associated density matrix is diagonal in spin space,
we can represent the non-interacting state $|\Phi_0^\alpha\rangle$ as
\begin{eqnarray}
|\Phi_0^\alpha\rangle &=& |\varphi_\uparrow^\alpha\rangle \otimes 
|\varphi_\downarrow^\alpha\rangle \\
|\varphi_\sigma^\alpha\rangle &=& a_{1,\sigma}^{\alpha,\dagger} 
a_{2,\sigma}^{\alpha,\dagger}
a_{3,\sigma}^{\alpha,\dagger} \dots a_{N\sigma,\sigma}^{\alpha,\dagger}
|0\rangle
\end{eqnarray}
and the operators $a_{k,\sigma}^\alpha$ are related to the real space
operators $c_{i,\sigma}$ by the linear transformation
\begin{equation}
a_{k,\sigma}^\alpha = \sum_i \phi_{i,\sigma}^\alpha(k) c_{i,\sigma}
\end{equation}
which defines the specific inhomogeneous solution. Details for the
calculation of the amplitudes $\phi_{i,\sigma}^\alpha(k)$ within the
Gutzwiller approximation can be found in Ref. \onlinecite{sei98}.
Within these definitions the evaluation of matrix elements between
different Slater determinants is analogous to the scheme outlined
in Ref. \onlinecite{louis99}.
Here we have defined the single-particle matrix elements as
\begin{eqnarray}
\langle k^\alpha_\sigma|q^\beta_\sigma\rangle &=& \sum_i
\phi_{i,\sigma}^\alpha(k)\phi_{i,\sigma}^\beta(q) \\
\langle k^\alpha_\sigma|n_{i,\sigma}|q^\beta_\sigma\rangle &=& 
\phi_{i,\sigma}^\alpha(k)\phi_{i,\sigma}^\beta(q) .
\end{eqnarray}

The matrix elements between Slater determinant and also those 
of single particle operators between
different Slater determinants as used e.g. in Eq. (\ref{eq:dble2})
are given by

\begin{equation}
\langle\Phi_0^\alpha|\Phi_0^\beta\rangle_\sigma= \left| \begin{array}{llcl}
\langle 1^\alpha_\sigma|1^\beta_\sigma\rangle &
\langle 1^\alpha_\sigma|2^\beta_\sigma\rangle &
\cdots &
\langle 1^\alpha_\sigma|N_\sigma^\beta\rangle \\
\langle 2^\alpha_\sigma|1^\beta_\sigma\rangle &
\langle 2^\alpha_\sigma|2^\beta_\sigma\rangle &
\cdots &
\langle 2^\alpha_\sigma|N_\sigma^\beta\rangle \\
\multicolumn{4}{c}\dotfill\\
\langle N^\alpha_\sigma|1^\beta_\sigma\rangle &
\langle N^\alpha_\sigma|2^\beta_\sigma\rangle &
\cdots &
\langle N^\alpha_\sigma|N_\sigma^\beta\rangle 
\end{array}\right| \label{eq:over}
\end{equation}

\newpage

\begin{eqnarray}\label{eq:det1}
\left\lbrack \hat{n}_{i,\sigma}\right\rbrack_{\alpha\beta}
&=& \left| \begin{array}{llcl}
\langle 1^\alpha_\sigma|n_{i,\sigma}|1^\beta_\sigma\rangle &
\langle 1^\alpha_\sigma|2^\beta_\sigma\rangle &
\cdots &
\langle 1^\alpha_\sigma|N_\sigma^\beta\rangle \\
\langle 2^\alpha_\sigma|n_{i,\sigma}|1^\beta_\sigma\rangle &
\langle 2^\alpha_\sigma|2^\beta_\sigma\rangle &
\cdots &
\langle 2^\alpha_\sigma|N_\sigma^\beta\rangle \\
\multicolumn{4}{c}\dotfill\\
\langle N^\alpha_\sigma|n_{i,\sigma}|1^\beta_\sigma\rangle &
\langle N^\alpha_\sigma|2^\beta_\sigma\rangle &
\cdots &
\langle N^\alpha_\sigma|N_\sigma^\beta\rangle 
\end{array}\right|  \\
&& \nonumber \\
&+& \,\,\,\,\, \left| \begin{array}{llcl}
\langle 1^\alpha_\sigma|1^\beta_\sigma\rangle &
\langle 1^\alpha_\sigma|n_{i,\sigma}|2^\beta_\sigma\rangle &
\cdots &
\langle 1^\alpha_\sigma|N_\sigma^\beta\rangle \\
\langle 2^\alpha_\sigma|1^\beta_\sigma\rangle &
\langle 2^\alpha_\sigma|n_{i,\sigma}|2^\beta_\sigma\rangle &
\cdots &
\langle 2^\alpha_\sigma|N_\sigma^\beta\rangle \\
\multicolumn{4}{c}\dotfill\\
\langle N^\alpha_\sigma|1^\beta_\sigma\rangle &
\langle N^\alpha_\sigma|n_{i,\sigma}|2^\beta_\sigma\rangle &
\cdots &
\langle N^\alpha_\sigma|N_\sigma^\beta\rangle 
\end{array}\right| +  \nonumber \\
&& \nonumber \\
&+& \cdots +
\left| \begin{array}{llcl}
\langle 1^\alpha_\sigma|1^\beta_\sigma\rangle &
\langle 1^\alpha_\sigma|2^\beta_\sigma\rangle &
\cdots &
\langle 1^\alpha_\sigma|n_{i,\sigma}|N_\sigma^\beta\rangle \\
\langle 2^\alpha_\sigma|1^\beta_\sigma\rangle &
\langle 2^\alpha_\sigma|2^\beta_\sigma\rangle &
\cdots &
\langle 2^\alpha_\sigma|n_{i,\sigma}|N^\beta_\sigma\rangle \\
\multicolumn{4}{c}\dotfill\\
\langle N^\alpha_\sigma|1_\sigma^\beta\rangle &
\langle N^\alpha_\sigma|2^\beta_\sigma\rangle &
\cdots &
\langle N^\alpha_\sigma|n_{i,\sigma}|N^\beta_\sigma\rangle 
\end{array}\right| \nonumber .
\end{eqnarray}

\end{document}